[1994/12/01]
\documentclass[cpp,fleqn]{w-art}
\usepackage{times}
\usepackage{amsmath}
\usepackage{w-thm}
\usepackage{bm}
\usepackage{graphicx}
\usepackage{epstopdf}

\chardef\bslash=`\\ 

\newcommand{\eval}[2][\right]{\relax
  \ifx#1\right\relax \left.\fi#2#1\rvert}

\begin{document}
\Volume{XX}
\Issue{1}
\Month{1}
\Year{2003}
\pagespan{1}{}
\Receiveddate{}
\Reviseddate{}
\Accepteddate{}
\Dateposted{}




\title[ Two-$T$ WDM pair potentials and phonon spectra ]{Two-temperature pair potentials and phonon spectra for simple metals in the warm dense matter regime}


\author[Harbour]{L. Harbour\footnote{{\sf Louis.Harbour@umontreal.ca},}\inst{1}} \address[\inst{1}]{D\'eŽpartement de physique et Regroupement qu\'eŽbŽcois sur les mat\'eŽriaux de pointe (RQMP), Universit\'eŽ de Montr\'eŽal, Montr\'eŽal, Qu\'eŽbec, Canada }

\author[Dharma-wardana]{M. W. C. Dharma-wardana\inst{2}}
\address[\inst{2}]{National Research Council of Canada, Ottawa, Ontario, Canada}

\author[Klug]{D. D. Klug\inst{2}}

\author[Lewis]{L. J. Lewis\inst{1}}


\begin{abstract}
We develop ion-ion pair potentials for Al, Na and K for densities and temperatures relevant to the warm-dense-matter (WDM) regime. Furthermore, we emphasize non-equilibrium states where the ion temperature $T_i$  differs from the electron temperature $T_e$. This work focuses mainly on ultra-fast laser-metal interactions where the energy of the laser is almost exclusively transferred to the electron sub-system over femtosecond time scales. This results in a  two-temperature system with $T_e>T_i$ and with the ions still at the initial room temperature $T_i=T_r$. First-principles calculations, such as density functional theory (DFT) or quantum Monte Carlo, are as yet not fully feasible for WDM conditions due to lack of finite-$T$ features, e.g. pseudopotentials, and extensive CPU time requirements. Simpler methods are needed to study these highly complex systems. We propose to use two-temperature pair potentials $U_{ii}(r, T_i,T_e)$ constructed from linear-response theory  using the non-linear electron density $n(\mathbf{r})$ obtained from finite-$T$ DFT with a {\it single} ion immersed in the appropriate electron fluid. 
We compute equilibrium phonon spectra at $T_r$ which are found to be in very good agreement with experiments. This gives credibility to our non-equilibrium phonon dispersion relations which are important in determining thermophysical properties, stability, energy-relaxation mechanisms and transport coefficients.

\end{abstract}

\noindent \{ For the refereed version, see: DOI http://dx.doi.org/10.1002/ctpp.201400092 \} \\ 
\{  International Conf. on Strongly-Coupled Coulombo Systems (SCCS) 2014 \} \\

\maketitle






\section{Introduction}

Warm dense matter (WDM) is a region of 
the phase diagram located at the conjuncture of the plasma state and the condensed-matter state. The density is high enough that ions are strongly correlated and can be found in a mixture of different ionization states. The ion temperature $T_i$ can be different from the electron temperature $T_e$ which can be as high as the Fermi temperature or even more, and hence the electrons can be partially degenerate. Such WDM states are of great interest at the fundamental level, but also for their technological applications such as laser ablation \cite{bib1}, inertial-confinement fusion \cite{bib2}, Coulomb explosion \cite{bib3}, etc., and are also of relevance to astrophysics \cite{bib4,chen2013}. It is a great challenge to predict  the two-temperature quasi-thermodynamic properties or transport and radiative processes of such basically non-thermodynamic states. The interest for WDM in the theoretical physics community has greatly increased since the advent of the possibility of creating these extreme conditions in the laboratory. However, the experimental techniques required for this purpose have the crucial particularity of always producing two-temperature WDM with $T_i\neq T_e$, which is problematic in the current theoretical framework. The present work focuses on the case where $T_e>T_i$, which can be observed when a femtosecond laser pulse interacts with a thin metal foil \cite{bib5, bib6}. Since the electron-ion mass ratio is very small, the laser energy is almost totally transferred to the electron system, thus increasing the electron temperature from the equilibrium initial temperature to $T_e$. Meanwhile, since the transfer of energy from the laser to the electron subsystem, as well as the equilibration of the latter, happens much faster than the electron-ion energy relaxation, the ion temperature $T_i$ remains at room temperature $T_r$ with the  ionic-lattice structure staying intact as the ions have no time to move. Using a two-temperature model (TTM) \cite{Milchberg88,bib7} to describe the electron-ion energy relaxation, it has been determined that the electron-ion relaxation time $\tau_{ei}$ is of the order of picoseconds, while $\tau_{ee}$ and $\tau_{ii}$ are orders of magnitude shorter. The challenge then is to develop theoretical tools designed for such non-equilibrium and strongly-correlated systems.

Within the framework of density functional theory (DFT) \cite{bib8}, the total free energy of a system of electrons and ions  $F[n(\mathbf{r}),\rho (\mathbf{r})]$ can be written as a functional of the one-electron density $n(\mathbf{r})$ and the one-ion density $\rho(\mathbf{r})$. The energy minimization results in a set of two coupled equations for $n(\mathbf{r})$ and $\rho (\mathbf{r})$. The first set is the integral equations for the ion distribution $\rho(\mathbf{r})$ which obeys classical physics, and hence can be solved  using the hypernetted-chain (HNC) approximation within the Ornstein-Zernike (OZ) equation if  the ion-ion pair potential $U_{ii}(\mathbf{r})$ is known. The second set is the Kohn-Sham \cite{bib9} equations for electrons in the external potential caused by $\rho(\mathbf{r})$. The HNC approximation, or its modified version with bridge corrections, provides the correlation potentials for the ion subsystem which does not have exchange effects. However, a key element in the DFT-equations of the electron subsystem is the exchange and correlation (xc) free energy functional $F_{xc}[n,T_e]$
and its density derivative which is the xc-potential. Major DFT codes --- e.g. ABINIT \cite{bib19} --- are designed for ground-state calculations using ground-state adapted pseudopotentials and do not include many finite-$T$ features. They offer a large variety of ground state xc-functionals $E_{xc}[n]$ but do not currently implement finite-$T$ xc-functionals. Since DFT calculations can be quite sensitive to the choice of $E_{xc}[n]$ in certain regimes, it seems very important to use a finite-$T$ functional $F_{xc}[n,T_e]$ to adequately study WDM systems. Furthermore, in the DFT extension to finite-$T$ by Mermin \cite{bib10}, it is required to include the Fermi distribution  $f(E_i,T_e)$ into the finite-$T$ electron density $n(r,T_e)$ calculation. Since $T_e$ can be very high in the WDM regime, it requires many excited Kohn-Sham-Mermin eigenfunctions in each part of the self-consistent procedure which considerably increases
the calculation time. Moreover, standard DFT simulations are designed for  crystal or molecular structures, which excessively limit the range of $T_i$ accessible to study WDM. 

Higher ion temperatures can be studied using molecular dynamics (MD) simulations. MD is a powerful method to study the ion system in the liquid, gaseous or plasma phases. It has been highly parallelized to make it possible to simulate very large systems. The main input of MD simulations is the ion-ion pair potential which is required for each interaction in the system.  However, most pair potentials, as implemented in comprehensive MD packages such as LAMMPS \cite{lammps}, are semi-empirical constructions. A particular example is the embedded-atom method (EAM) \cite{bib16}, fitted to reproduce particular properties under ambient conditions while failing at predicting other ones, especially those of interest in two-temperature WDM. 

In the following, we propose to construct ion-ion two-temperature  pair potentials (TTP) $U_{ii}(r,T_i,T_e)$ based on fundamental considerations. The first part of this paper explains how we construct the non-equilibrium electron-ion pseudopotentials $U_{ei}(r,r_s,T_{ei})$, which 
depend on both temperature $T_{ei}\equiv (T_e, T_i)$ and density via the electron Wigner-Seitz radius $r_{s}$. The most important quantity to compute is the free electron density $n_f(r,T_{ei})$ around an ion of effective charge $\bar{Z}$. This density is then used to construct the pseudopotential from linear-response theory, going beyond the random phase approximation (RPA) by adding a finite-$T$ local-field correction (LFC). The second part explains how to construct the pair potential $U_{ii}$ from $U_{ei}$ and shows the importance of the LFC. The total ion-ion interaction is obtained by adding the direct-ion Coulomb interaction to the indirect interaction through the electron subsystem. In the last part, we compute the equilibrium and non-equilibrium phonon spectra $\omega_\mu(k,T_{ei})$ of three simple metals to test our TTP $U_{ii}(r,T_{ei})$. We examine aluminum as well as sodium and potassium which are typical free-electron metals under the conditions studied.

\section{Non-equilibrium electron-ion pseudopotentials}

According to the original formulation of DFT by Hohenberg and Kohn \cite{bib8}, the electron density $n(\mathbf{r})$ rather than the many-body wavefunction is considered to be the fundamental quantity describing a quantum system so that {\it all} physical properties of interest can be expressed as a functional of $n(\mathbf{r})$. The self-consistent scheme proposed by Kohn and Sham \cite{bib9} enables one to reduce the many-body system to an effective ``one-electron'' problem by putting all the missing information about the electron-electron interactions in the xc-functional $E_{xc}[n]$. However, since electrons in the WDM regime are far from zero temperature, it is crucial to use the extension of DFT to finite temperatures. Furthermore, one could wonder if using the DFT for systems with $T_i \neq T_e$ is in contravention with the fundamentals of DFT, but since  calculations are done for a single fixed ion surrounded by its $T_i$-dependent density which is immersed in an electron liquid at $T_e$, the applicability of DFT is satisfied.

\subsection{Electron density from finite-$T$ DFT}

In extending the work of Hohenberg and Kohn to finite-$T$, Mermin  showed that the electron density defines uniquely the external potential $V(r)$ of the system \cite{bib10}. The Mermin-Kohn-Sham equations  reduce the finite-$T$ many-electron problem to a one-electron problem, giving the interacting  electron density $n(\mathbf{r},T_e)$.
\begin{equation}
n(\mathbf{r},T) = n_b(\mathbf{r}) +n_f(\mathbf{r}) = \sum_i^N f(\epsilon _i,T)|\phi_i(\mathbf{r})|^2,  \qquad \text{where} \qquad  f(\epsilon_i,T)=\frac{1}{1+\exp[\beta(\epsilon_i-\mu^0)]},
\end{equation}
with $n_b(\mathbf{r})$ and $n_f(\mathbf{r})$ the bound and free electron density,  $f(\epsilon_i)$ the Fermi occupation function, $\mu^0(T)$ the non-interacting chemical potential, $\beta=1/k_bT$ the inverse temperature (with $k_b=1$ in our units), and $\phi_i(\mathbf{r})$ the Mermin-Kohn-Sham wavefunction of eigenvalue $\epsilon_i$. At $T=0$, $\mu^0$ is the Fermi energy $\epsilon_F$. As mentioned in the introduction, the presence of the Fermi occupation factors, and the need for high angular-momentum states, greatly increase the CPU time since the convergence of the energy is quadratic in the number of bands needed. One way to decrease the simulation time and make it accessible to WDM conditions is to reduce the number of electrons in the system by introducing the effective ion charge $\bar{Z}$ which appears in the pseudopotential. However, the formulation of finite-$T$ pseudopotentials remains largely
unexplored, and standard codes simply use the zero-$T$ ones which are inadequate for many WDM applications~\cite{Trickey12}.

\subsubsection{Effective charge $\bar{Z}$ and the neutral pseudoatom approximation}

The main objective is to replace the nuclear charge $Z$ by an effective charge $\bar{Z}\equiv Z- n_b$ where $n_b$ is the number of bound electrons. This is calculated by integrating the bound electron density $n_b(\mathbf{r})$ over the ion sphere \cite{PerrotBe,MuriDW2013}. The main difficulty is to determine how to split the total charge density into a bound and a free part  $n(\mathbf{r})=n_b(\mathbf{r})+n_f(\mathbf{r})$ since  bound electrons may extend beyond the Wigner-Seitz radius $r_{s}$. In this study, there is no ambiguity in the value of $\bar{Z}$ for Al, Na and K in the given density and temperature range. Using $\bar{Z}$, one may attempt to construct  an electron-ion pseudopotential which reproduces the free electron wavefunctions outside a core radius $r_c$. However, a large number of continuum wavefunctions are needed in the WDM regime associated with finite-$T$ problems~\cite{Trickey12}, and a large  number of ions have to be included in the simulation box of the usual type of DFT calculations, increasing significantly the CPU time. We simplify the problem using two key ideas from DFT. (i) Only one nucleus at the origin is considered, and spherical symmetry can be exploited; hence $\mathbf{r}$ will be replaced by $r$ when appropriate. The charge density $n(r)$ around the central nucleus, rather than its wavefunctions are considered. (ii) Instead of including every ion at their locations $\mathbf{r}_i$ we consider only the one-ion charge distribution $\rho(\mathbf{r}) =\sum_i\delta(\mathbf{r}-\mathbf{r}_i)$.  Here we follow the correlation-sphere-DFT model of Dharma-wardana and Perrot and  compute the electron density around only one ion of effective charge $\bar{Z}$ surrounded by the self-consistently calculated spherically averaged ion density {\it distribution} $\rho(r)$\cite{bib11}.
Once the free electron charge density $n_f(r)$ at an ion in the given WDM environment is determined, a simple local pseudopotential can be extracted  as explained in sec.~\ref{lnearized-pseudo.sec}.

 For the elements that we study in this work, it is possible to simplify 
the calculation of $n_f(r)$ even more by using the neutral-pseudoatom (NPA) approximation~\cite{Dagens73}. The NPA approach is valid for solids, liquids and plasmas at low or high densities and temperatures as long as  $\bar{Z}$ can be clearly defined so that core-electrons are not important in the ion-ion interaction. Thus, the ion density is replaced by a uniform positive background (jellium) except for a spherical neutral cavity of radius $r_{\rm WS}$ (ion Wigner-Seitz radius) around the central ion. In this model, the ion temperature $T_i$ is merely a parameter which determines the value of $r_{\rm WS}$ from the density of the material at $T_i$. All these simplify greatly the solution of the Kohn-Sham-Mermin equations.

\subsubsection{The exchange and correlation free energy functional}

To compute $n(\mathbf{r})$ around the single ion, the finite-$T$ xc-potential given by Perrot and Dharma-wardana is used in the  Kohn-Sham-Mermin equations. It has been constructed as follows. The fully non-local xc free energy per electron $f_{xc}=F_{xc}/N$ is given by the integral of the spin-dependent electron-electron correlation function $h_{ij}(\mathbf{r})=g_{ij}(\mathbf{r})$-1  over the coupling constant $\lambda$ 
\begin{equation}
f_{xc}=\frac{n}{2}\int_0^1 \frac{d\lambda}{2\lambda}\int d\mathbf{r}\frac{\lambda}{r}[h_{\uparrow\uparrow}(\mathbf{r},\lambda)+h_{\uparrow\downarrow}(\mathbf{r},\lambda)].
\end{equation}
Using the classical-map hypernetted-chain (CHNC) \cite{bib12} method,  $h_{ij}(r)$ and $f_{xc}$ have been calculated by Perrot and Dharma-wardana for different electron densities $r_{s}$ and temperatures $T_e$. Their results  appear to be in good agreement  with the $T=0$ and finite-$T$ quantum Monte Carlo (QMC)  calculations \cite{bib13}, thereby supporting the use of the CHNC approximation for problems where QMC data are not available. The method consists in  mapping the quantum electron-fluid onto a classical Coulomb fluid using an effective electron-electron potential and an effective electron temperature to simulate quantum effects. It makes it possible to solve the OZ integral equations with the HNC approximation so as to get the finite$-T$ interacting $h_{ij}(\mathbf{r}, \lambda, r_s, T_e)$. The $f_{xc}(r_s,T_e)$ obtained this way has been parametrized to a simple form \cite{bib14} which makes it very practical.

\subsection{Electron-ion pseudopotentials from linear response theory}
\label{lnearized-pseudo.sec}
Once the effective charge $\bar{Z}$ and the free electron density $n_f(r)$ have been calculated for a particular $r_{s}$ and $T_{ei}$, it is possible to construct a weak local ($s-$wave) electron-ion pseudopotential $U_{ei}(r)$ using linear response theory \cite{bib15}. In  $q$-space, it takes the form
\begin{equation}
U_{ei}(q,r_s,T_{ei}) = n_f(q,r_s,T_{ei})/\chi_{ee}(q,r_s,T_e),
\end{equation}
with $\chi_{ee}$ the finite-$T$ interacting electron linear response. Only the $q<\sim 3k_F$ region is relevant as the ion-core is excluded.  To go beyond the RPA, we include a local-field correction $G_q$ so that $\chi_{ee}$ is given by
\begin{equation}
\chi_{ee}(q,r_s,T_e)=\frac{\chi_0(q)}{1-V_q(1-G_q)\chi_0(q)} \quad \text{with} \quad V_q =\frac{4\pi}{q^2}, \quad G_q=\left(1-\frac{\gamma_0}{\gamma} \right)\left(\frac{q}{k_{TF}} \right)^2.
\end{equation}
Here $\chi_0(q)$ is the non-interacting finite$-T$ Lindhard function which may use an effective electron mass $m^*$ if needed~\cite{bib15}. The LFC is used in the local-density approximation (LDA) and is evaluated from the ratio of the non-interacting and interacting finite$-T$ compressibilities $\gamma_0$ and $\gamma$ respectively of the electron subsystem. The interacting  compressibility can be evaluated from the finite-$T$ xc-potential. It is important to note that the NPA approach is valid if resulting pseudopotentials satisfy $U_{ei}(q)/(-\bar{Z}V_q) \leq 1$.

\section{Two-temperature ion-ion pair potentials}
\subsection{The ion-ion pair potentials}
It is now possible to construct two-temperature pair potentials by adding the direct ion-ion interaction to the indirect interaction through the electron fluid such that
\begin{equation}
U_{ii}(q,r_s, T_{ei})=\bar{Z}^2(T_i)V_q+|U_{ei}(q,r_s,T_{ei})|^2 \chi_{ee}(q,r_s,T_e).
\end{equation}
The extension to a mixture of ions of different types or different ionizations is discussed in Ref.~\cite{bib17}. In the present work, we present two-temperature pair potentials resulting from this procedure for three different simple metals. We examine aluminum, which is the ``standard'' free electrons metal with three valence electrons ($\bar{Z}=3$), followed by sodium and potassium, which are alkali metals with one valence electron ($\bar{Z}=1$). Since most  laser pulse-probe experiments are done at room temperature, we computed pair potentials with $T_i=T_r=0.026$ eV whereas $T_e$ is increased up to some value, e.g. 6 eV, that is experimentally relevant. The resulting TTP for Al, Na and K are presented in Fig. \ref{fig:1}. The pseudopotentials used to construct these TTP are fitted to a Heine-Abarankov form (HA). A simple discussion of HA may be found in Shaw and Harrison~\cite{HA-in-Harrison}. The
local HA pseudopotentials use a core radius $r_c$ and a constant core-potential $A$ for $r<r_c$, while the potential for $r>r_c$ is $\bar{Z}/r$. The equilibrium lattice parameter at room temperature $a_L$ resulting from such potentials is usually 1-3\% in error. We have slightly modified the $r_c$ and $ A$ parameters to obtain nearly the correct $a_L$. 
This is justified since the ion positions are held fixed at their room-temperature values in the WDM studied here.

\begin{vchfigure}[h]
\centering
\includegraphics[scale=0.54]{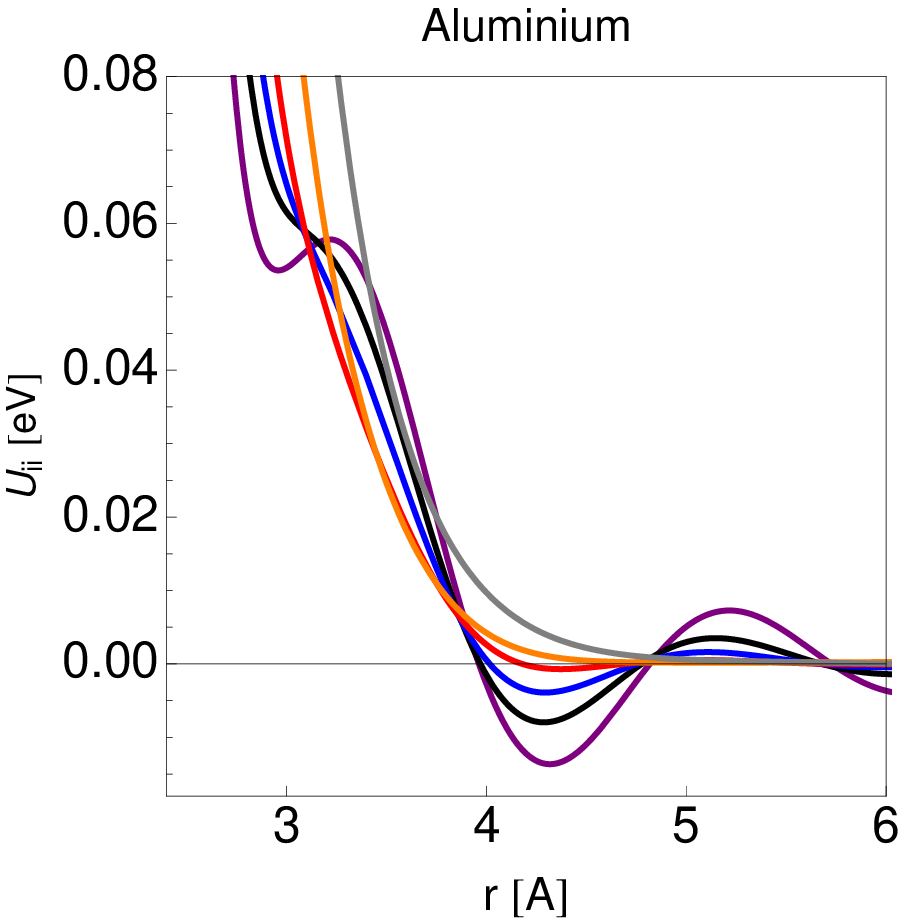}
\includegraphics[scale=0.47]{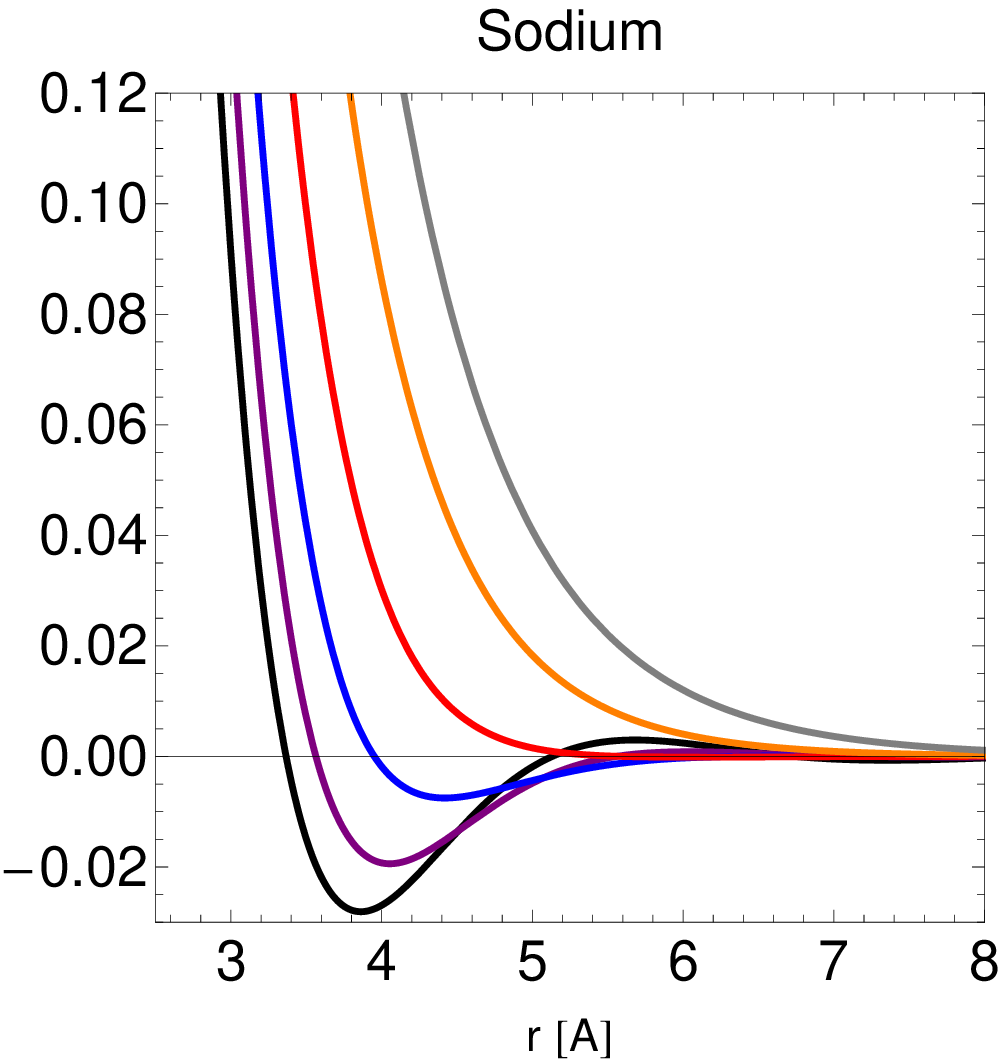}
\includegraphics[scale=0.475]{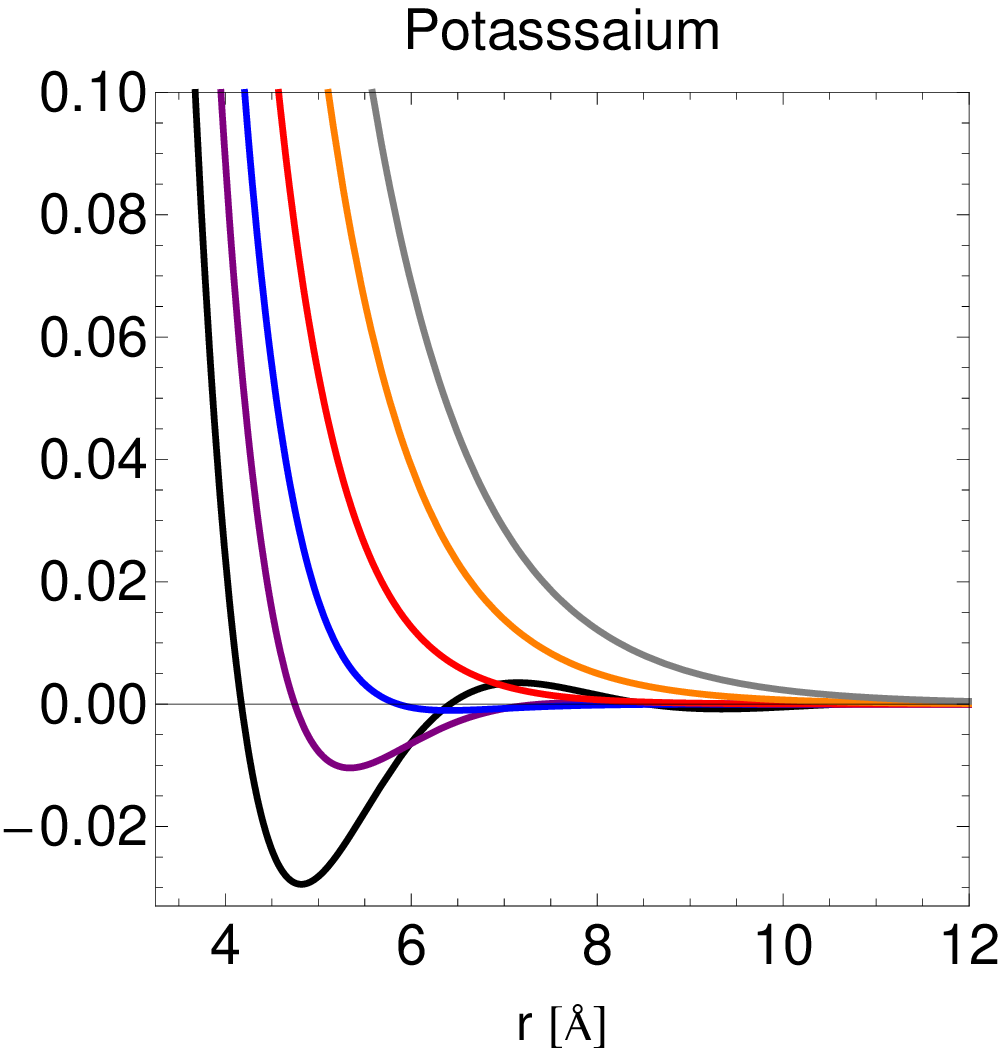}
\vchcaption{Two-temperature pair potentials for  Al, Na and K. Ions are at $T_i=0.026$ eV and the electrons at $T_e$ = 0.026 eV (black), 1.25 eV (purple), 2.00 eV (blue), 3.00 eV (red), 4.50 eV (orange) and 6.00 eV (gray). }
\label{fig:1}
\end{vchfigure}

An important characteristic of these pair potentials is the presence of long-range oscillations which result from the Friedel oscillations in the electron density. These are related to the discontinuity in the Fermi distributions in WDM and zero-$T$ systems.
They are clear non-linear effects included in our model from the self-consistently 
calculated electron density and exist even in linear response theory which is used to construct the pseudopotentials. Since even the first minimum is basically a Friedel oscillation, it is very important to accurately calculate their amplitude. It has been shown that fitting pair potentials to elastic constants or phonon dispersion curves results in non-unique amplitudes \cite{bib18}. We can see that the screening effect depends greatly on the number of valence electrons as the amplitudes of  oscillations in the Na and K potentials  damp much faster than in the Al case. Furthermore, as $T_e$ increases, the electron screening becomes less and less Friedel-like and the pair potentials exhibit  only the repulsive behavior typical of Yukawa potentials. At such $T_e$, the crystal structure usually looses its
stability, resulting in a Coulomb explosion or melting of the metal, depending on the heating time scales.

\subsection{The importance of the local-field correction}
To illustrate the importance of going beyond the RPA  form of the screening function $\chi$, we computed the pair potential for Al at $T_{ei}=T_r$ with and without the LFC; the results are presented in Fig.\ref{fig:2}.

\begin{vchfigure}[h]
\centering
\includegraphics[scale=0.65]{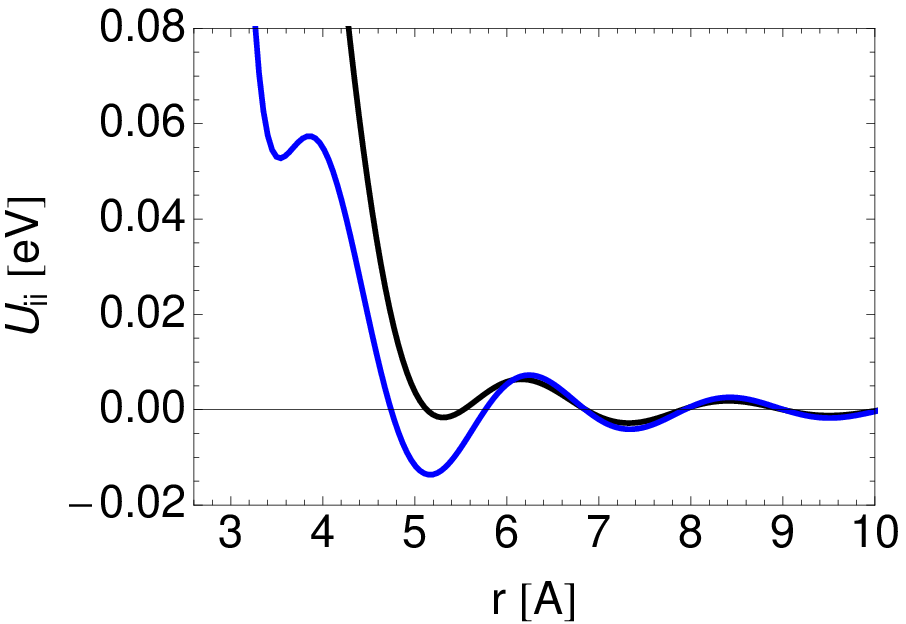}
\vchcaption{Comparison between Al pair potential at $T_{ei}=T_r$ obtained using the RPA (black line) and with the LFC  included (blue line) in the screening function.}
\label{fig:2}
\end{vchfigure}

We observe a notable difference between the two pair potentials. The  RPA-potential does not exhibit the first minimum shown by the  LFC potential. This difference is crucial since this first minimum defines the nearest-neighbour position in the crystal; thus the RPA does not give a realistic description of solid Al.

\section{Phonon spectrum of laser-heated simple metals}

Using the TTP, it is possible to compute the phonon energies by constructing the dynamical matrix~\cite{AshcroftMermin}:
\begin{equation}
\mathbf{D}(\mathbf{q})= \sum_{i}\mathbf{D}(\mathbf{r}_i)e^{-i\mathbf{k}\cdot \mathbf{r}_i} \qquad \text{with} \qquad \mathbf{D}(\mathbf{r_i}) =\left.\frac{\partial^2 U_{ii}(\mathbf{r}_i-\mathbf{0})}{\partial \mathbf{u}(\mathbf{r}_i)\ \partial\mathbf{u}(\mathbf{0})}\right|_{\mathbf{u}=0}.
\end{equation}
Here $\mathbf{u}$ is the small displacement in the position of  ions at $\mathbf{r}_i$ and at the origin $\mathbf{0}$. Phonon frequencies are then given by $\omega_\mu(\mathbf{q})=\sqrt{\lambda_\mu(\mathbf{q})/M}$, with  $\lambda_\mu(\mathbf{q})$ the eigenvalues of the dynamical matrix. Using pair potentials, it is extremely rapid to compute the phonon dispersions for any crystal structure. To validate our simple model, we compare it with a more fundamental linear-response method using the ABINIT package \cite{bib19}. The latter is, strictly speaking, only applicable to systems in equilibrium, with $T_i=T_e$. Determining phonon spectra in DFT is usually done using the response function \cite{bib201,bib202} method where  the first derivative of every wavefunction is computed.  This requires even more CPU time. Typically, while the ABINIT phonon calculation for aluminum may take many hours on a computer cluster, the pair potential approach is nearly `instantaneous' on a laptop.

The ions in the ultra-fast laser-heated solid  are not at their equilibrium positions but frozen in their `old' positions. The phonon description assumes that the elementary excitations are oscillations of the ions about their equilibrium positions, with the first derivative of the potential, i.e., the force, equal to zero. This is increasingly invalid as $T_e$ exceeds $T_i$. However, this `harmonic approximation' is assumed in both the ABINIT and the pair-potential phonon calculation given here.  In reality, the first derivative (i.e., the Hellman-Feynman force on each ion) is large and will lead to an explosive acceleration of the ions outwards rather than generating harmonic oscillations. A consideration of the time scale for one oscillation, and the time scale for the motion of an ion by about a wavelength, suggest that only the higher frequency modes can have a physical meaning.  Nevertheless, we present the full phonon spectra in order to compare with similar efforts of other workers (using {\it ab initio} codes) who have provisionally assumed the existence of these non-equilibrium phonons. Besides the first-derivative being negligible, the second derivative of the total potential, being the elastic constant for the harmonic motion, must be positive at the ion positions. The higher derivatives give anharmonic effects. Thus, at sufficiently high $T_e$, the  phonon frequencies  become  purely imaginary, as the pair-potential becomes more like a Yukawa potential. It is extremely important to experimentally determine `phonon spectra'  of WDM systems to establish the scope and validity of the phonon description of the elementary excitations of these two-temperature solids. 

\subsection{Aluminum}
At room temperature, aluminum exhibits a FCC crystal structure with an experimentally determined lattice parameter $a_{\text{Al}}=4.05 \textup{\AA}$ which is used as an input (via the density) into our TTP construction. On the other hand, our $\textit{ab initio}$ simulations are done using the LDA with the norm-conserving pseudopotential method. We used a 165 eV kinetic energy cutoff and a $20\times 20\times 20$ Monkhorst-Pack $\mathbf{k}$-point grid to compute the planewave expansion of the pseudowavefunctions. 
The ground-state calculation gives us an equilibrium lattice parameter $a_{\text{Al}}^{\text{DFT}}=4.01 \textup{\AA} $ and phonon frequencies, at $T_{ei}=T_r=0.026$ eV, in very good agreement with our TTP model and with experiment \cite{bib21}, as presented in Fig. 3. We computed the phonon spectra for the laser-heated situation where ions are kept `frozen' at their initial equilibrium position, and with $T_i=T_r$,  whereas 
the electron temperature is increased up to $T_e=6$ eV. The non-equilibrium phonon spectra obtained with our TTP model and with ABINIT-DFT, by simulating more electronic band and using 
 the Fermi-Dirac occupation function to mimic finite-$T$ effects , are also displayed in Fig. 3. 

\begin{figure}[h]
\hfill
\begin{minipage}{72mm}
\includegraphics[scale=0.695]{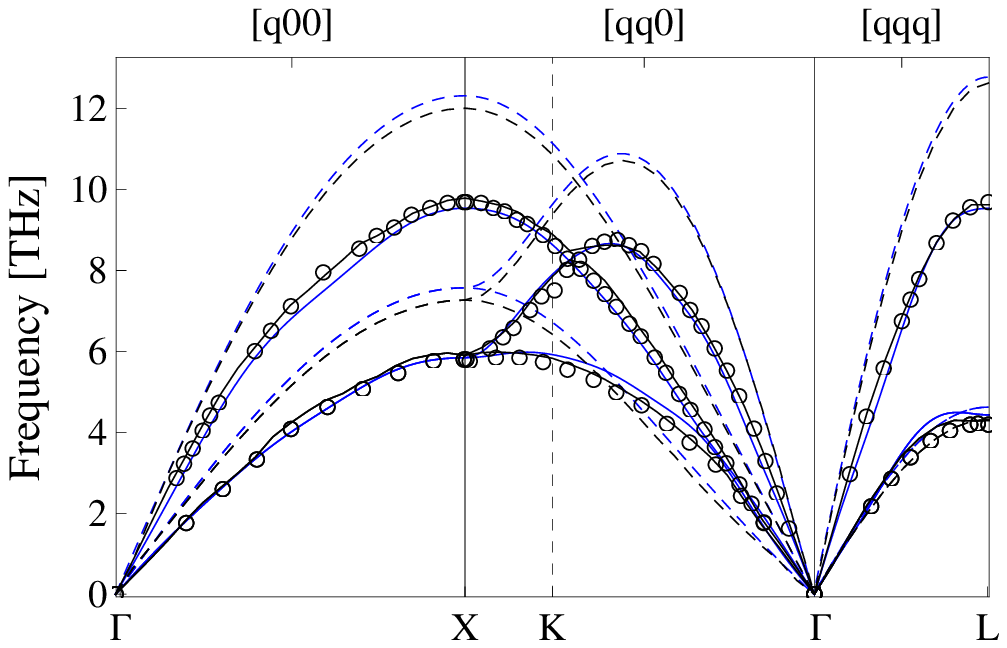}
\caption{Equilibrium (full lines) and non-equilibrium (dashed lines) aluminum phonon dispersion relations computed with the TTP model (blue) and ABINIT (black), compared with experimental data (circles).}
\label{fig:3}
\end{minipage}
\hspace{10mm}
\begin{minipage}{72mm}
\vspace{2mm}
\includegraphics[scale=0.62]{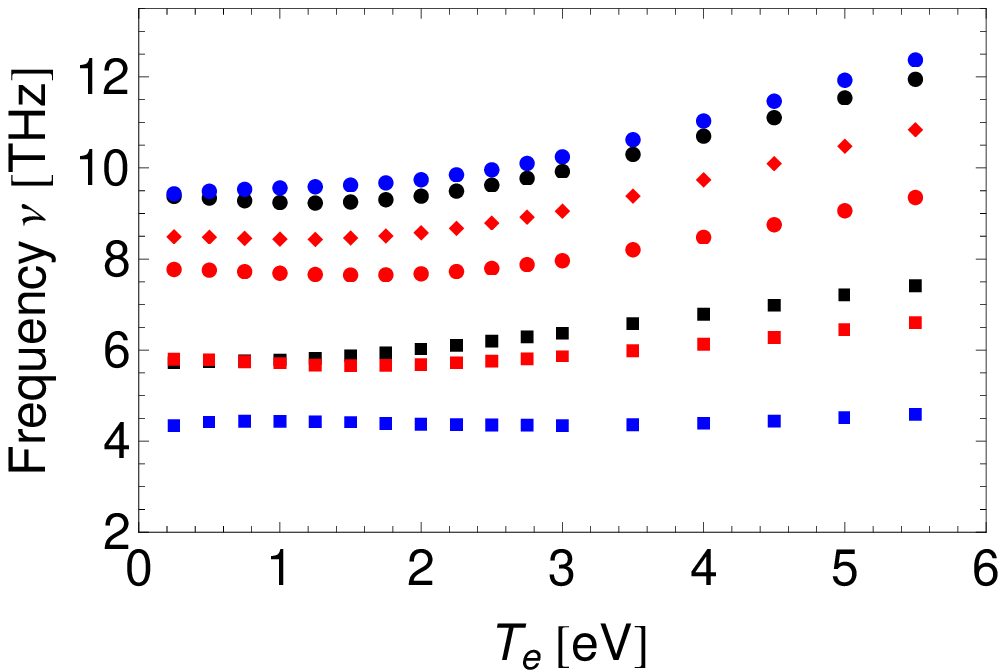}
\caption{Phonon frequencies of laser-heated aluminum at $T_i=T_r\neq T_e$ for longitudinal (circles) and transverse (squares and diamonds) modes at symmetry points X (black), L (blue) and K (red).}
\label{fig:4}
\end{minipage}
\end{figure}

As $T_e$ increases, we find that both methods show phonon hardening. The TTP model predicts a maximum frequency increase  of 29$\%$ at the symmetry point X passing from 9.54 THz to 12.32 THz for the longitudinal mode. As mentioned by Recoules \textit{et al}  \cite{bib22}, the transverse mode frequency at the symmetry point L is not really affected, passing from 4.43 THz to 4.62 THz. Nevertheless, our results clearly illustrate that all other branches are also modified by the laser-heating mechanism. At room temperature, the TTP model predicts frequencies lower than DFT calculations whereas at $T_e=6$ eV they are higher. As $T_e$ increase, finite$-T$ effects become more and more important; further studies are needed to understand if this difference comes from the finite$-T$ features which are absent in the ABINIT-DFT simulations. The fast computation of TTP by this method permits us to compute phonons for many different $T_e$ as presented in Fig.4. The ability to calculate phonon spectra `on the fly' can be of great utility in determining electron-ion energy relaxation and non-equilibrium transport properties such as the thermal and electrical conductivity.

\subsection{Sodium and potassium}
Sodium and potassium are alkali metals having a BCC crystal structure at room temperature with experimentally determined lattice parameters of $a_{\text{Na}}=4.23 \textup{\AA}$ and $a_{\text{K}}=5.23 \textup{\AA}$, respectively, which will be used as input in the TTP construction. The ABINIT-DFT simulations have been done within the generalized-gradient approximation (GGA), a $24\times 24\times24$ Monkhorst-Pack $\mathbf{k}$-point grid and a 1360 eV kinetic energy cutoff for both Na and K. Ground state calculations result in lattice parameter $a_{\text{Na}}^{\text{DFT}}=4.20 \textup{\AA}$ and $a_{\text{K}}^{\text{DFT}}=5.25 \textup{\AA}$, respectively, and yield phonon frequencies in agreement with our TTP model and with experiment \cite{bib23, bib24} as presented at Fig.5. 

\begin{figure}[h]
\includegraphics[scale=0.745]{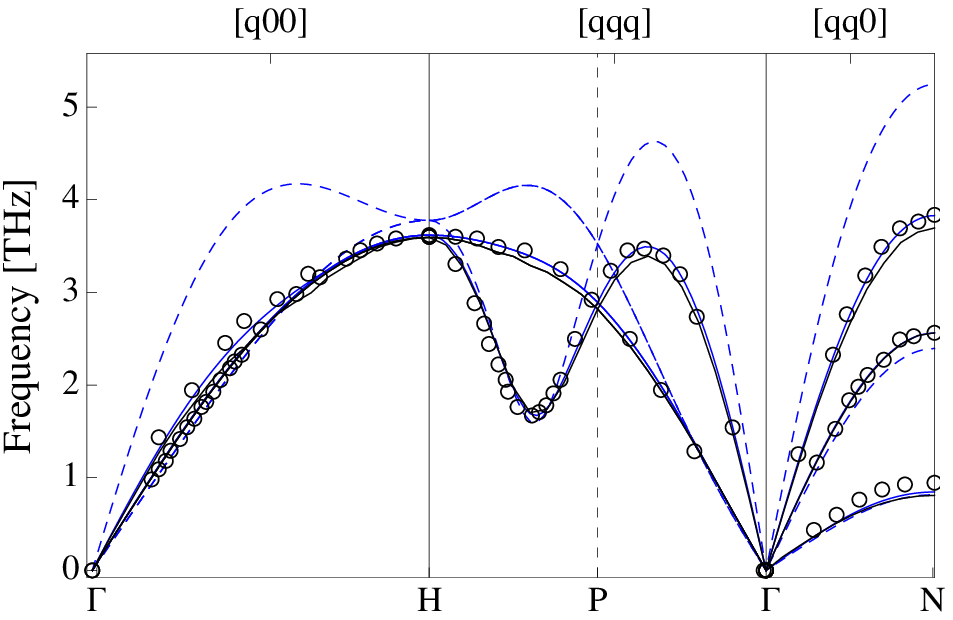}\ a)
\hfil
\includegraphics[scale=0.78]{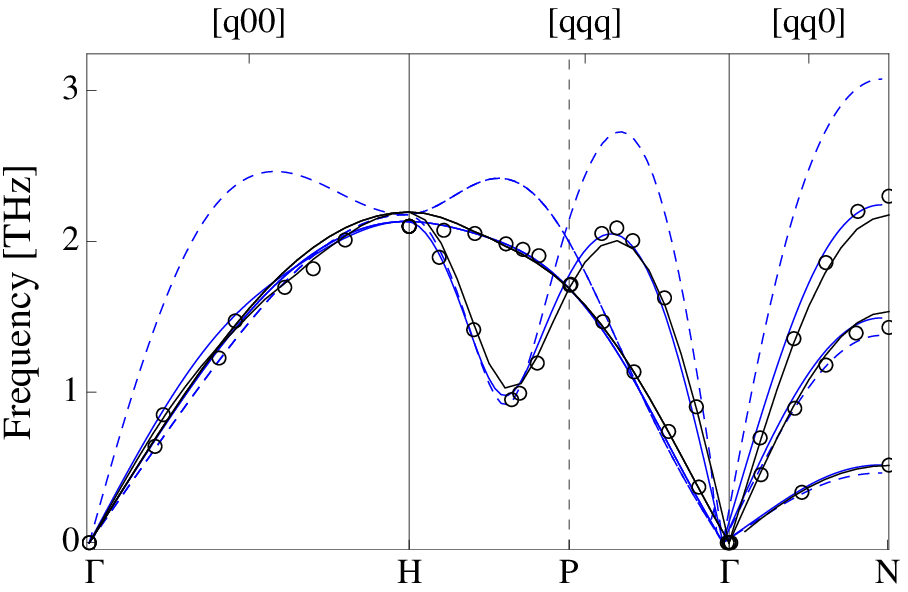}\ b)
\caption{Equilibrium $T_e=T_i=T_r$ (full lines) and non-equilibrium $T_e =6$ eV (dashed lines) phonon dispersions  for $\mathbf{a)}$ sodium and $\mathbf{b)}$ potassium computed with the TTP model (blue) and ABINIT (black), compared with experiment (circles).}
\label{fig:5}
\end{figure}

Once again, our TTP results are in excellent accord with experiment for the equilibrium case which gives us confidence in  applying the method to the non-equilibrium situation. In the case of Na and K, we observe that most frequencies increase but also that the shape of the dispersion relation changes as $T_e$ reaches 6 eV. We can see  new maxima appearing between symmetry points $\Gamma- H$ and between $H-P$. In both cases, the maximum increase occurs at the point N with an augmentation for the frequency of the longitudinal branch of 37$\%$. Finite$-T$ \textit{ab initio} simulations for these metals are still not completed. Hence the comparison of the  predictions of our TTP model with more fundamental calculations is not yet final for this case.

\section{Conclusion}
As WDM studies are becoming more and more topical and important in many fields of physics, we have addressed the problem of  describing theoretically such highly-correlated systems via a simple model based on transparent physics. Furthermore, to compare predictions with experiment, it is necessary to describe non-equilibrium systems where $T_i\neq T_e$. Since \textit{ab initio} simulations can be very time consuming in this regime, we have developed simple two-temperature pair potentials as the `work-horse' of the method. They are constructed from  fundamental considerations by computing the electron density from finite-$T$ DFT and then using linear response theory with a finite-$T$ local-field correction. Pair potentials resulting from this procedure produce long-range interactions resulting from the presence of Friedel oscillations in the electron density. These TTP were used to compute phonon dispersion relations for equilibrium Al, Na and K. The excellent accord with experiment and DFT simulations (using the ABINIT code) establishes support for the validity of our approach for the non-equilibrium case. In all cases, our model predicts an increase of the frequencies, which is different for each mode branch, supporting the phonon hardening hypothesis. At higher $T_e$, where thermal effects are more important, frequencies obtained for Al with the TTP model are found to be higher than the ones obtained by DFT simulations. This discrepancy might come from the non-implementation of a finite-$T$ exchange and correlation energy functional and the lack of other finite-$T$ features in  \textit{ab initio} codes like  the ABINIT. To validate this hypothesis, further work will be done to implement our finite$-T$ xc-functional in a major DFT code. The possibility to quickly compute TTP for many different combinations of $T_i$ and $T_e$ could also be used to predict electron-ion energy relaxation and non-equilibrium transport properties. Another important application of our TTP is to incorporate it in  laser ablation simulations where pair potentials dependent on both temperatures are needed. Finally, construction of two-temperature potentials for more complex metals, such as gold and copper, would be of great interest since many WDM experiments \cite{bib25} have been carried out with these metals.

\section{Acknowledgments}

This work has been supported by grants from the Canadian Natural Sciences and Engineering Research Council (NSERC) and the Fonds de Recherche Qu\'ebec - Nature et Technologies
(FQRNT). We are indebted to Calcul Qu\'{e}bec for generous allocations of computer resources.

\end{document}